\documentclass[aps,showpacs,floatfix,twocolumn]{revtex4}
\usepackage{graphicx}
\usepackage{amssymb}
\newcommand{\be}{\begin{eqnarray}}
\newcommand{\ee}{\end{eqnarray}}
\def\beq{\begin{equation}}
\def\eeq{\end{equation}}

\begin{document}
\title{A holographic approach to phase transitions}
\author{Sebasti\'an Franco}
\affiliation{KITP, University of California, Santa Barbara, CA93106-4030, USA}
\author{Antonio M. Garc\'{\i}a-Garc\'{\i}a}
\affiliation{CFIF, IST, Universidade T\'{e}cnica de Lisboa, Av. Rovisco Pais, 1049-001 Lisboa, Portugal}
\affiliation{Physics Department, Princeton University, Princeton,
New Jersey 08544 , USA}
\author{Diego Rodr\'{\i}guez-G\'{o}mez}
\affiliation{Queen Mary, University of London, Mile End Road, London E1 4NS, UK}
\begin{abstract}
We provide a description of phase transitions at finite temperature in strongly coupled field theories using holography. For this purpose, we introduce a general class of gravity duals to superconducting theories that exhibit various types of phase transitions (first or second order with both mean and non-mean field behavior) as parameters in their Lagrangian are changed. Moreover the size and strength of the conductivity coherence peak can also be controlled. Our results suggest that certain parameters in the
gravitational dual control the interactions responsible for binding the condensate and the magnitude of its fluctuations close to the transition.
\end{abstract}
\maketitle
Second order phase transitions are one of the more intensively investigated phenomena in theoretical physics.
Part of this interest stems from the fact that the dynamics of a system close to a second order phase transition is, to a large extent, universal.
Observables around the transition are
controlled by scaling variables which depend on a few independent critical exponents. Mean field theories provide an adequate theoretical
framework for describing the transition only if fluctuations of the order parameter can be neglected. This typically occurs only for high dimensional systems.
 In lower spatial $d \leq 2$ dimensions, rigorous results are only known for some systems \cite{spherical}.
 In the rest of cases the effect of fluctuations is accounted by extrapolating small $\epsilon$ expansions to $\epsilon = 1$ or assuming that the number of components of the condensate is large.

In this paper we provide a description of phase transitions in strongly interacting systems
 using the AdS/CFT correspondence \cite{adscft}. This correspondence provides a theoretical framework to describe strongly coupled conformal field theories (CFTs) through a weakly coupled dual gravitational description. The relevance of the AdS/CFT correspondence in this context comes from the fact that, close to second order phase transitions, the correlation length diverges and the system is well approximated by an interacting CFT. In this work, we take a phenomenological approach and focus directly on a bulk gravity
theory designed to capture universal properties of the putative
dual field theory. 
Since we want to model a conformal field theory by using the AdS/CFT correspondence, we assume this
gravity theory lives in an assymptotically AdS space.
Temperature is added by introducing a black hole in the AdS geometry. By assumption, the CFT of interest undergoes a phase transition in which a global continuous symmetry is spontaneously broken at low temperatures by the vacuum expectation value (VEV) of a charged scalar field. In the relevant subsector of the gravitational counterpart this corresponds to a bulk gauge field dual to the global current and a scalar field, both propagating in the bulk asymptotically AdS geometry. The interactions with the gravity background must be such that the global symmetry of the condensate is spontaneously broken at sufficiently low temperatures. This follows from the generation of a non-trivial hair profile for the black hole; which is dual to the boundary VEV triggering the symmetry breaking. Gravity models with this property \cite{Gubser:2008px} have attracted considerable interest \cite{Hartnoll:2008vx,Gubser:2008zu} for their potential applications to condensed matter physics \cite{reviews}. Our aim is to investigate a large class of phase transitions (that is, classes of CFT's) for which our phenomenological perspective is applicable.

We will restrict ourselves to systems in two spatial dimensions and condensates with a global $U(1)$ symmetry which is spontaneously broken by a generalization of the St\"uckelberg mechanism \cite{stu38}.
The models we introduce could in principle be extended to other spatial dimensions and to condensates with other symmetries, such as $SU(N)$. Such generalizations might shed light on the dynamics of phase transitions in field theories with spontaneous chiral symmetry breaking. Moreover, since the dynamics close to a second order transition is to a great extent universal, our findings might provide an effective description of second order phase transitions in strongly interacting condensed matter systems.

The main finding of this letter is that indeed the broader setting provided by a generalized St\"uckelberg mechanism \cite{stu38} of symmetry breaking allows for a description of a fairly wide class of phase transitions. This framework allows tuning the order of the phase transition and, for second order phase transitions, the value of critical exponents. Indeed, in a certain range of parameters, the conductivity in our model is similar to that observed in some strongly coupled superconductors \cite{hightc,huse1} close to the superconductor-metal transition.

\begin{figure}[ht!]
\includegraphics[width=0.72\columnwidth,clip,angle=0]{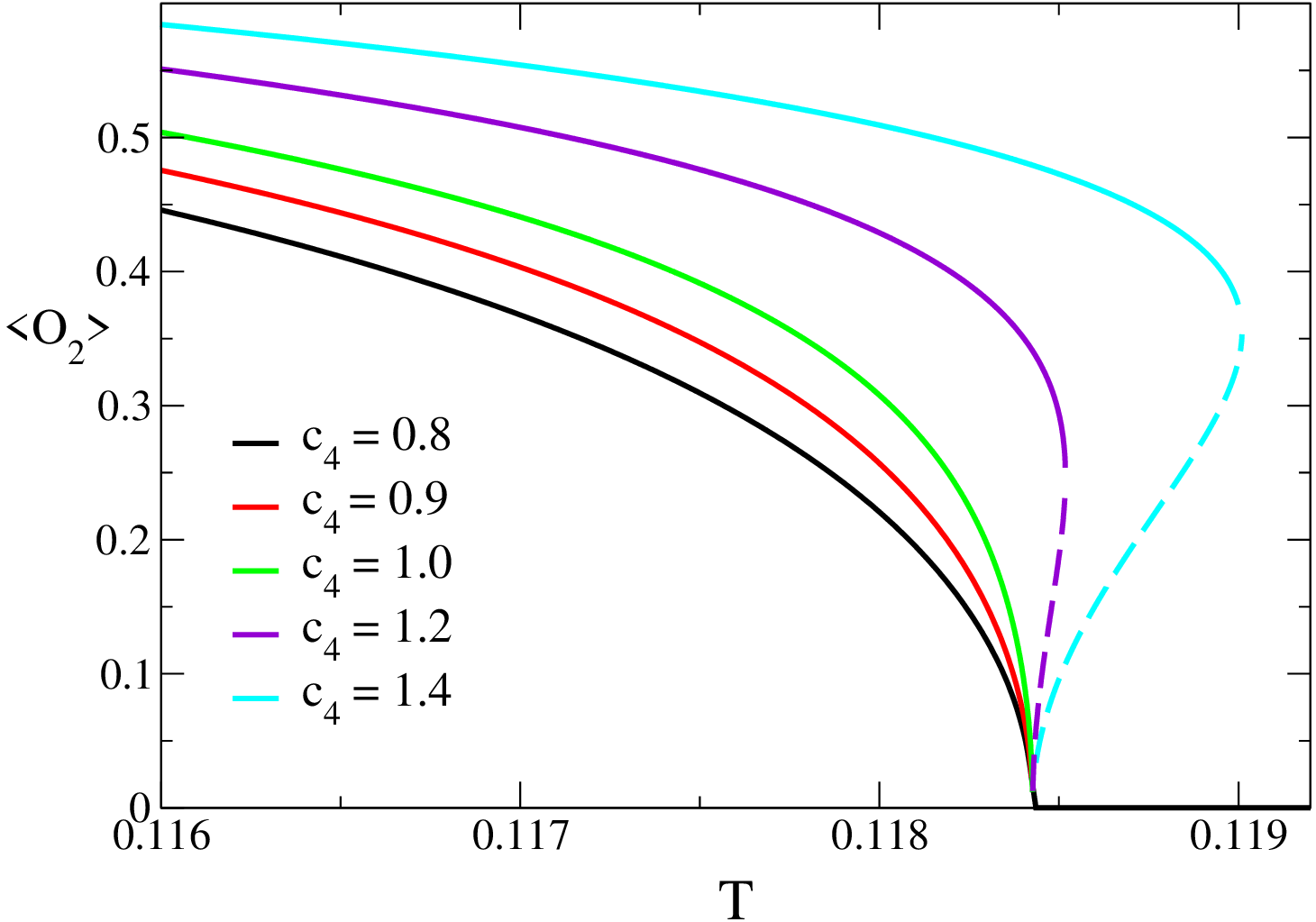}
\includegraphics[width=0.72\columnwidth,clip,angle=0]{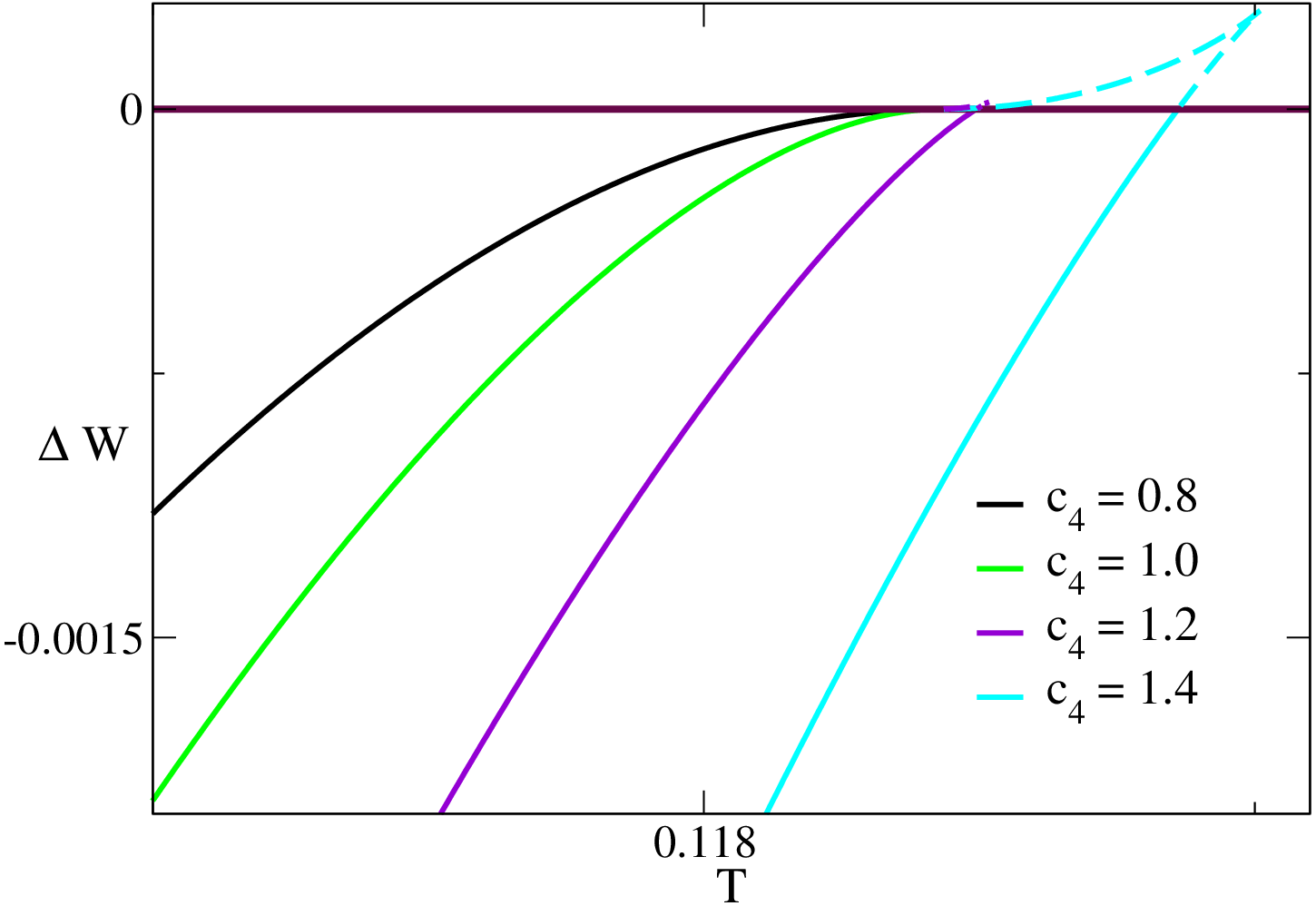}
\caption{Upper: Normalized condensate $\langle\mathcal{O}_2\rangle$ as a function of temperature at fixed $\rho = 1$ for $\mathcal F$ given by (\ref{f1}) with $c_3 = 0$, and different $c_4$'s. A jump in the condensate at the critical temperature is clearly observed for $c_4 \gtrapprox 1$. Dashed lines delimit the metastable region typical in first order phase transitions. Lower: Free energy difference  $\Delta W(T) = W - W_0$ with $W$ given by Eq. (\ref{W1}), $W_0 = 1/2r_H$ the free energy in the uncondensed phase, $c_3 = 0$, and different $c_4$'s. 
$\Delta W$ is not analytical at $T_c$ ($\Delta W(T_c) = 0$) for $c_4 \gtrapprox 1$. This is a signature of a first order phase transition.}
\label{O1fullCanonicalEnsemble}
\end{figure}

Returning to our phenomenological approach, one of the simplest models with the required features consists of a $U(1)$ gauge field ($A_0=\Phi$) and real scalars ($\tilde\Psi$ and $p$), coupled via a generalized St\"uckelberg Lagrangian (see \cite{us} for more details),
\begin{equation}
\label{gmodel}
S=\int\,\sqrt{g}\,\Big\{ -\frac{F^2}{4}-\frac{\partial\tilde{\Psi}^2}{2}-\frac{m^2\,\tilde{\Psi}^2}{2}-\left|\frac{\mathcal{F}(\tilde{\Psi})}{2}\right|\big(\partial p-A)^2\Big\}
\end{equation}
in an (asymptotically) AdS$_{d+2}$ background (dual to a CFT in $d+1$ spacetime dimensions) with a black hole. $\mathcal{F}$ is a general function. We note that for $\mathcal{F}={\tilde{\Psi}^2}$ it reduces to the Abelian Higgs model considered in \cite{Gubser:2008px,Hartnoll:2008vx,Maeda:2009wv}. In that case the phase transition is of second order with mean field critical exponents \cite{Maeda:2009wv}.
In what follows we restrict to the case of
$d = 2$ spatial dimensions.
Condensation of $\tilde\Psi$ at some temperature leads to spontaneous breaking of the $U(1)$ symmetry. While in (\ref{gmodel}) we have included an absolute value for $\mathcal{F}$ to avoid a wrong sign for the kinetic term, it is straightforward to find functions $\mathcal{F}$ that are positive definite for every $\tilde{\Psi}$. In general, it is possible to rewrite (\ref{gmodel}) (modulo high order terms in $\mathcal{F}$ which have a small effect in the dynamics) in terms of a complex scalar $\Psi$ of modulus $\tilde{\Psi}$ and phase $p$ such that $\tilde{\Psi}$ is constrained to be positive. We will use this property later. Further potential terms for $\tilde\Psi$, in addition to the mass term, can in principle be added to (\ref{gmodel}), but have little effect on the phase transition, since it is controlled by the black hole horizon. We work in the probe limit, in which gravity is decoupled from the rest of fields \cite{Hartnoll:2008vx}. This is a good approximation since, close to the transition, the magnitude of the scalar field is small and consequently its backreaction on the metric is expected to be negligible.

The geometry of the AdS$_{4}$ black hole takes the form
\begin{equation}
ds^2=-f(r)\,dt^2+\frac{dr^2}{f(r)}+r^2\, d\vec{x}_{2}^2\ ,~f(r)=r^2-\frac{M}{r}
\end{equation}
where $M$ is the mass of the black hole, $T = 3M^{1/3}/4\pi$ is its Hawking temperature and the AdS radius is $L = 1$. We use the gauge freedom to fix $p = 0$. We study the dynamics of $\Psi \equiv {\tilde \Psi}$ and $\Phi$ in this background, where $\Phi$ is the time component of the gauge field (the other components are turned off for the moment). We will also assume that both $\Psi$ and $\Phi$ are only functions of $r$. With these simplifications, the equations of motion become

{\small
\begin{eqnarray}
\label{bb}
\Psi''-\frac{z^3+2}{z\,(1-z^3)}\Psi'-\frac{m^2}{z^2\,(1-z^3)}\,\Psi+\frac{1}{r_H^2}\,\frac{\mathcal{F}'(\Psi)\Phi^2}{2(1-z^3)^2} = 0\\
\Phi''-\frac{\mathcal{F}(\Psi)}{z^2\,(1-z^3)}\,\Phi = 0 \nonumber
\end{eqnarray}}
\noindent where $z=\frac{r_H}{r}$, and $r_H=M^{1/3}$ is the horizon radius.
In these coordinates, the horizon sits at $z=1$ while the boundary of the AdS space is at $z = 0$.
Close to the boundary, we have

\begin{eqnarray}
\label{cc}
\Psi\sim\Psi_+\,r_H^{\lambda_+}\,z^{\lambda_+}+\Psi_-\,r_H^{\lambda_-}\,z^{\lambda_-} \\
\Phi \sim \mu-\frac{\rho}{r_H^{d-2}}\,z^{d-2} \nonumber
\end{eqnarray}
with $\lambda_{\pm}=(3\pm\sqrt{9+4\,m^2})/2$, and $\mu$ and $\rho$ a chemical potential and a charge density, respectively. For concreteness we assume that $m^2= -2$  (other values of $m^2$ are expected to give qualitatively similar results \cite{Horowitz:2008bn}). In this case, $\Psi\sim \frac{\Psi_1}{r_H}\, z+\frac{\Psi_2}{r_H^2}\, z^2\,+\ldots$ as $z \to 1$. The scalar condensate is given by $\langle\mathcal{O}_i\rangle=\Psi_i$, $i=1,2$. At the horizon, $\Phi$ vanishes due to the normalizability of $\Phi dt$ and $\Psi$ is regular. For the moment, we consider $\mathcal{F}$ of the simple form
\begin{equation}
\label{f1}
\mathcal{F} =\Psi^2 + c_3\Psi^3 + c_4\Psi^4,
\end{equation}
with $c_3$ and $c_4$ real numbers. As was already explained for $c_3 = c_4 = 0$ it reduces to the model of \cite{Gubser:2008px,Hartnoll:2008vx,Gubser:2008zu}.
The dual CFT interpretation of the theory with $\mathcal{F} \approx \Psi^2 + \Psi^4/3$ has been recently discussed \cite{gub2009}.
We will further consider positive $c_4$ and constraint $c_3$ such that $\mathcal{F}$ is positive definite and we can drop the absolute value in (\ref{gmodel}). It is important to emphasize that a polynomial choice of $\mathcal{F}$ does not indicate that our approach to the phase transition is {\it a la} Landau-Ginzburg. Even though a complete understanding is still lacking, it is reasonable to conjecture that the choice of $\mathcal{F}$ is linked to the properties of the corresponding CFT (or a subsector of it). From the effective theory point of view that we are taking, changing $\mathcal{F}$ corresponds to a sort of ``non normalizable deformation", and as such it corresponds to a change of theory. This might come either from an actual change of the microscopic theory or from a different truncation of it. This is in part justified by the findings in \cite{gub2009}.

The free energy is given by the on-shell action. Its calculation via a gravity dual is explained in detail in \cite{Hartnoll:2008kx,us}. The final result is
\begin{equation}
\label{W1}
W =-\frac{\mu\, \rho}{2}+\frac{r_H}{4}\,\int_0^1dz\, \frac{{\Psi\mathcal F'}(\Psi)\Phi^2}{z^2(1-z^3)}.
\end{equation}
\begin{figure}[!ht]
\includegraphics[width=0.72\columnwidth,clip,angle=0]{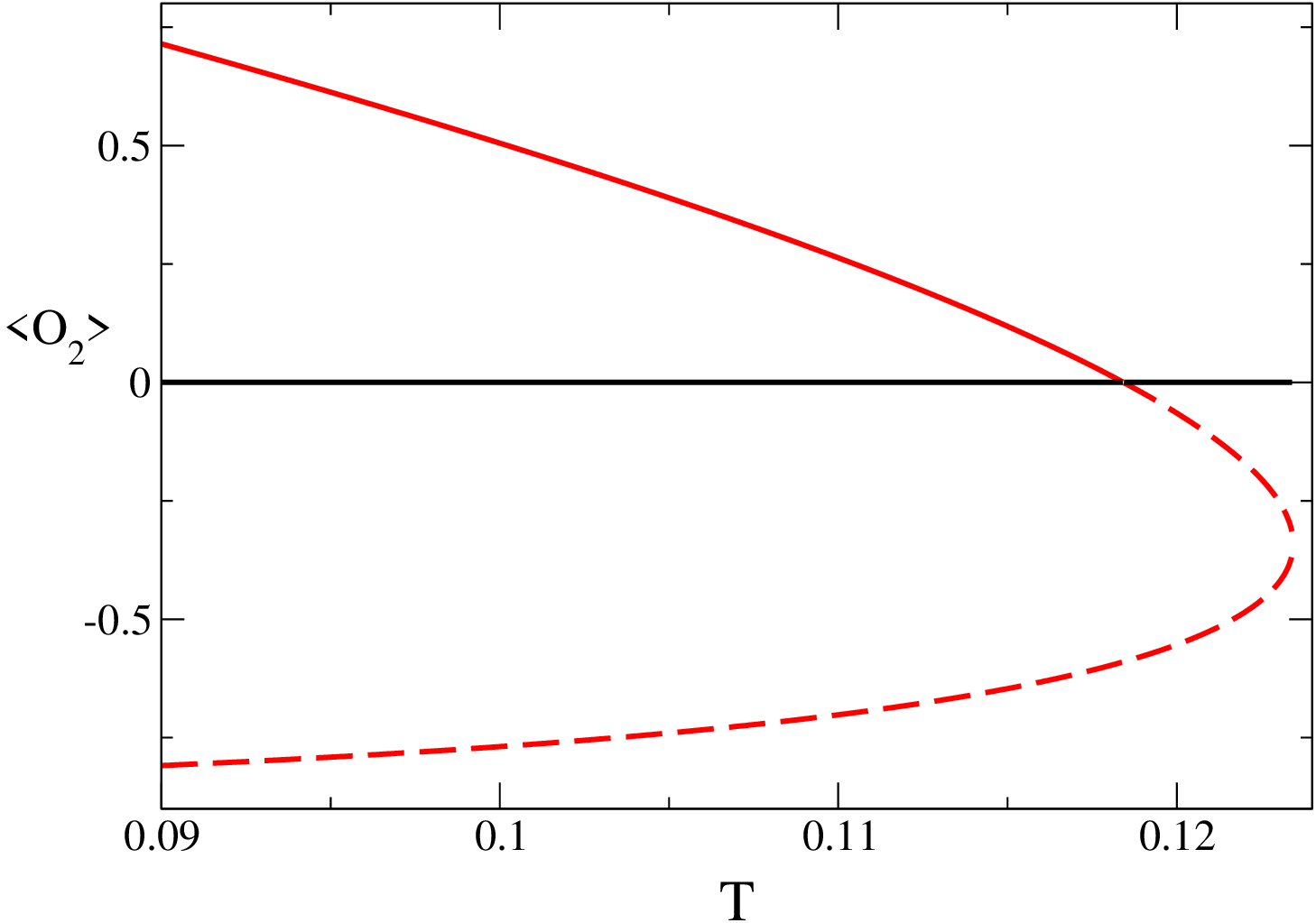}
\caption{$\langle \mathcal{O}_2\rangle$ condensate as a function of temperature at fixed $\rho = 1$ for $\mathcal F$ given by (\ref{f1}) with $c_3 = -1$ and $c_4 = 1/2$.}
\label{negative_c3}
\end{figure}
Let us now investigate how the phase transition depends on the coefficients $c_3$ and $c_4$, solving the equations of motion (\ref{bb}) numerically. In figure 1. (Upper), we plot the condensate around the critical region for $c_3=0$, $\rho =1$ and different values of $c_4$. For $0\leq c_4 \lesssim 1$, the transition is second order and the condensate approaches zero as $\mathcal{O}_i(T)\sim (T_c-T)^\beta$, with mean field critical exponent $\beta=1/2$. For $c_4 \gtrapprox 1$, the condensate does not drop to zero continuously at the critical temperature. While we have focused on the $\mathcal{O}_2$ condensate, analogous results are obtained for $\mathcal{O}_1$, with a different value of $c_4$ separating the first and second order behavior. Figure 1. (Lower) shows the free energy as a function of temperature. For $c_4 \gtrapprox 1$, the free energy develops a singularity at the critical temperature (at which the free energy of the condensed phase becomes equal to the one for the non-condensed phase). Both behaviors indicate that the phase transition changes from second to first order at $c_4 \approx 1$. Similarly, we have observed the phase transition becomes first order for $c_3 > 0$ (not shown).
\begin{figure}[!ht]
\centering
\includegraphics[width=0.72\columnwidth,clip,angle=0]{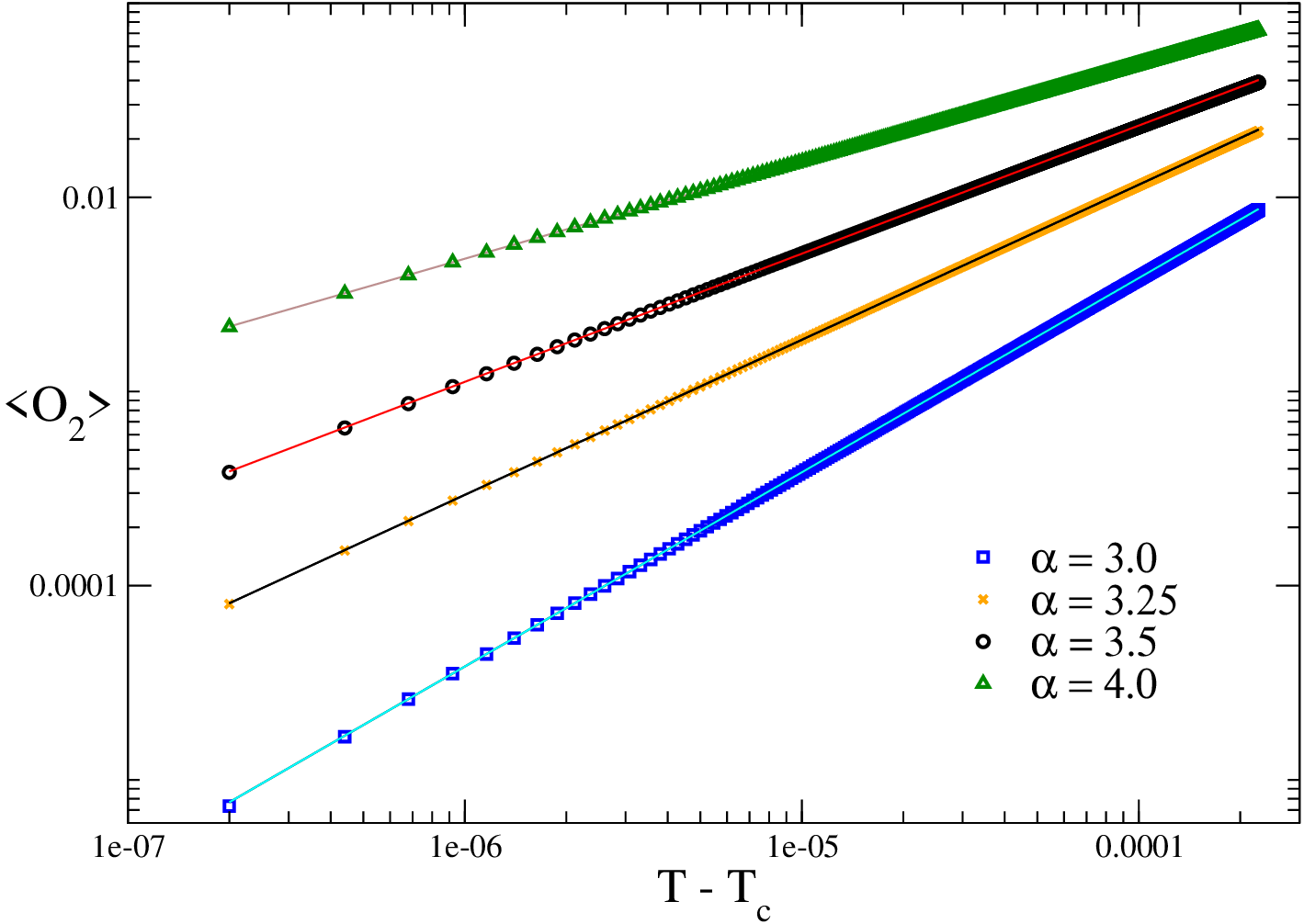}
\caption{Condensate in the proximities of $T_c$ for $c_\alpha = -1$ and different values of the exponent $\alpha$ in (\ref{falpha}). The critical exponent $\beta$ (the slope of the curves) clearly depends on $\alpha$. The best linear fit yields $\beta \approx 1, 0.80, 0.65, 0.5$ for $\alpha = 3, 3.25, 3.5, 4$ respectively. In fact, solving the equations of motion in power series around the horizon, it is possible to show that the critical exponent $\beta$ is related to $\alpha$ by $\beta = (\alpha-2)^{-1}$.
}
\label{loggap}
\end{figure}
\begin{figure}[!ht]
\centering
\includegraphics[width=0.72\columnwidth,clip,angle=0]{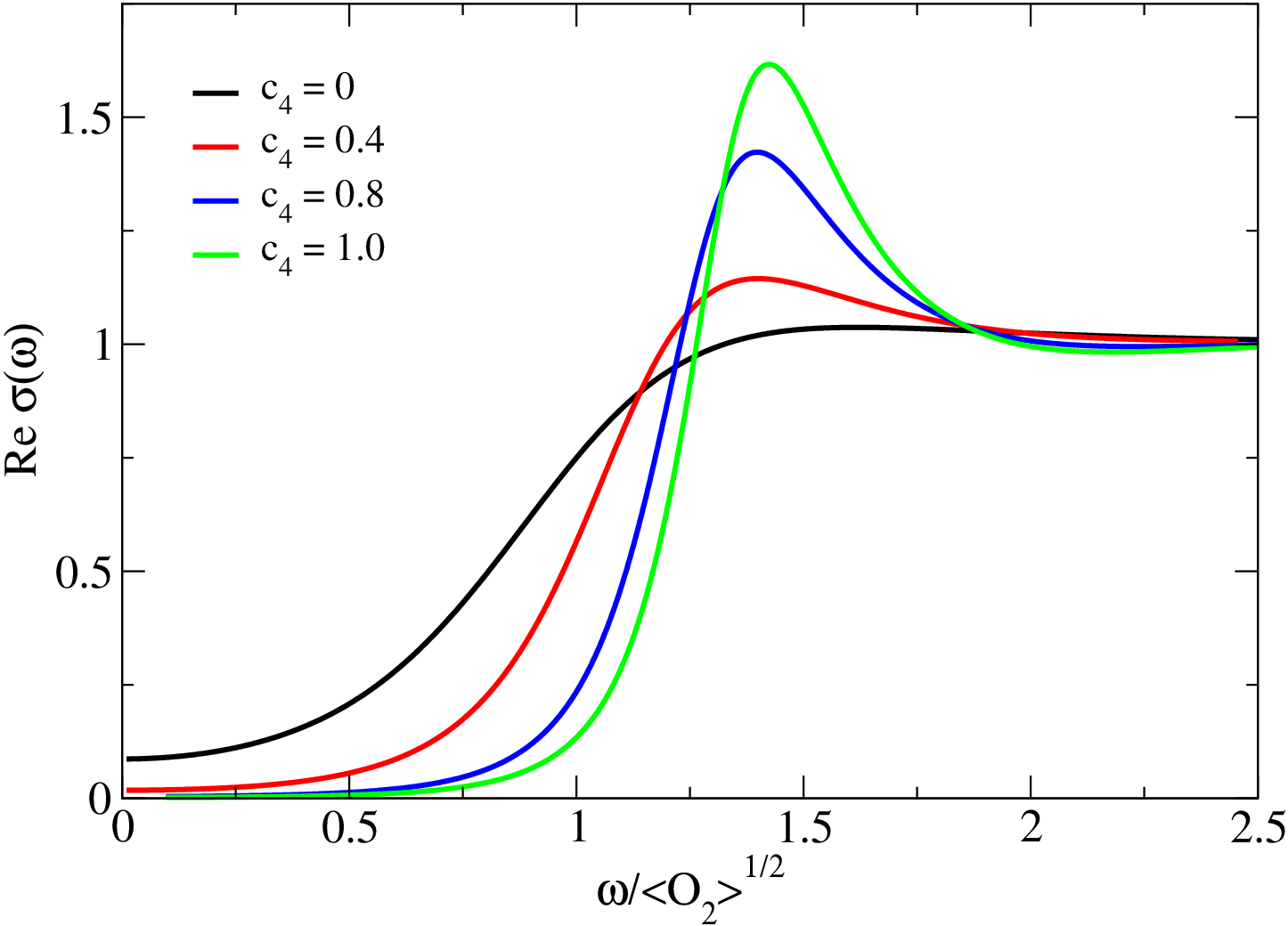}
\includegraphics[width=0.72\columnwidth,clip,angle=0]{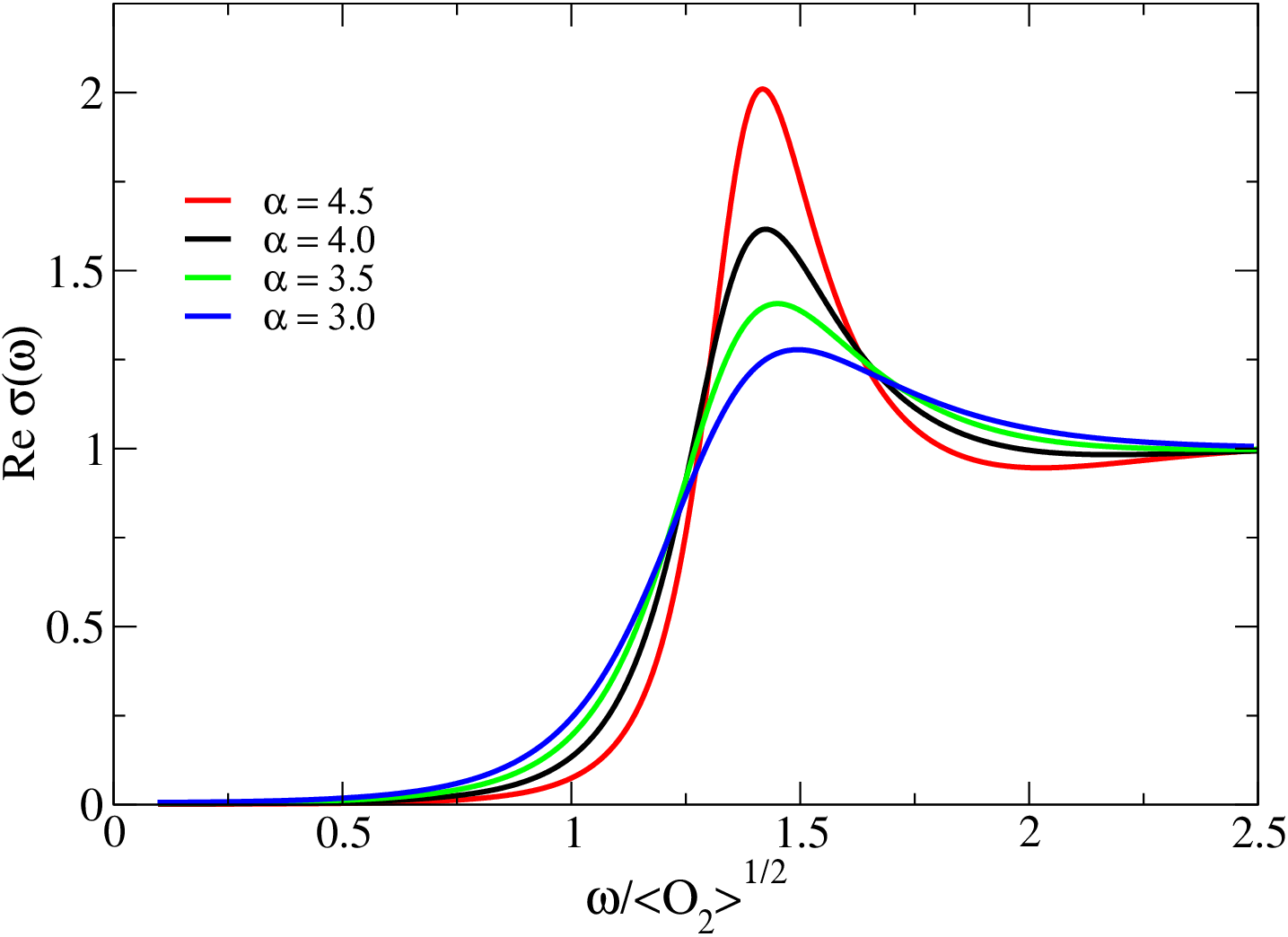}
\caption{Upper: ${\rm Re}[\sigma]$ (\ref{condu}) as a function of $\omega$ for $T \sim 0.84\, T_c$, $c_3=0$ and different values of $c_4$.
We observe the spectroscopic gap and the coherence peak that becomes narrower as $c_4$ increases. These features are similar to those of certain strongly coupled superconductors \cite{huse1,hightc}. Lower: A similar plot for $c_\alpha = 1$ and different values of $\alpha$ in (\ref{falpha}).
The coherence peak is more pronounced as $\alpha$ increases. This is an indication that condensate fluctuations also increase with $\alpha$.}
\label{conductivity}
\end{figure}

Having shown that gravity duals can lead to both first and second order phase transitions, it is natural to explore whether it is possible to find other choices of $c_3$ and $c_4$ such that the transition is of second order but with critical exponents different from the mean field prediction.
For $c_3 = c_4 = 0$ it has been recently shown  \cite{Maeda:2009wv} that critical exponents agree with the mean field values. We note that, though there are exceptions \cite{aiz}, in low dimensions the critical exponents of real systems deviate from the mean field prediction (see chapter $1$ of \cite{spherical} for a concise review). An interesting behavior arises for $c_3 < 0$. Figure 2 presents the condensate as a function of temperature for $c_3 = -1$ and $c_4 = 1/2$. We have included negative condensates for clarity. As explained earlier, this model can be rewritten in terms of a complex field in such a way that $\Psi$ is constrained to be positive. With this restriction, the phase transition is of second order (see figure 3) with $\beta = 1$. Such non-mean field behavior arises in this case as the result of a vertical shifting of the condensate curve towards negative values and restricting it to the positive branch.
More general (see figure 3)  critical exponents $1/2 \leq \beta \leq 1$ can be obtained by considering
\begin{equation}
\mathcal{F} =\Psi^2 + c_\alpha\Psi^\alpha+ c_4 \Psi^4,
\label{falpha}
\end{equation}
with $3 \leq \alpha \leq 4$. 
 An exponent $\beta$ larger than the mean field predictions indicates that fluctuations are suppressed and the condensate is particularly stable. This has been observed in the Gross-Neveu model for massless fermions  \cite{neveu:2001}.
 Such feature might then be an indication that interactions in the dual conformal field theory are long range and the system has chiral symmetry \cite{chi}. Percolation is other example in which exponents $\beta > 1/2$ have been found \cite{spherical}.

Finally, let us study the conductivity. In order to calculate it, one considers an electromagnetic perturbation $A_x = A_x(r)e^{-i\omega t + ik y}$ in the bulk.
The equation of motion for $A_x$  \cite{Hartnoll:2008kx,Horowitz:2008bn,reviews} is
\begin{equation}
\label{eq:Axeq}
A_x'' + \frac{f'}{f} A_x' + \left(\frac{\omega^2}{f^2}+\frac{k^2}{r^2f} - \frac{
{\mathcal F}(\Psi)}{f} \right) A_x = 0 \,.
\end{equation}
We must impose in-going wave boundary conditions at the horizon for causality \cite{Son:2002sd}:  $A_x  \propto f^{-i\omega/3r_0}$. At the boundary,
$
A_x = A^{0} + \frac{A^{1}}{r} + \cdots
$. The retarded Green function $G_R(\omega,k)$ is then given by
$G_R = A^1/A^0$. Finally, the conductivity is
\begin{equation}
\label{condu}
\sigma(\omega)=\frac{G_R(\omega,0)}{i\, \omega}.
\end{equation}
In figure 4. (Upper) we plot ${\mathcal Re} \, \sigma(\omega)$ for
the $\langle\mathcal{O}_2\rangle$ quantization at $T \sim 0.84\, T_c$, $c_3 = 0$ and different $c_4$'s.  The plots clearly show the existence of a gap that increases with $c_4$. We also observe a coherence peak that gradually becomes narrower and stronger as $c_4$ increases. In strongly interacting condensed matter systems, this type of behavior is caused by fluctuations of the condensate \cite{huse1,hightc}. This indicates that $c_4$ effectively controls the magnitude of the fluctuations. In fact, we have already seen that for $c_4 \gtrsim 1$ the fluctuations are strong enough to even induce a first order transition. Figure 4. (Lower) shows a similar plot but keeping $c_\alpha = 1$ fixed in (\ref{falpha}) and varying the exponent $\alpha$. It is observed that the coherence peak becomes more pronounced as we increase $\alpha$. Therefore the parameter $\alpha$ also controls the strength of fluctuations in the system.

In conclusion, we have introduced a class of gravity duals that is sufficiently general to describe both first and second order phase transitions at finite temperature in strongly interacting systems. For second order transitions, critical exponents different from the mean field prediction are obtained for some range of parameters in our model.
The size and strength of the coherence peak can also be controlled by the parameters that define $\mathcal F$. This is an indication
that our model may provide an effective description of certain aspects of strongly interacting systems close to $T_c$ with $\mathcal F$ controlling features such as the magnitude of the interactions binding the condensate or the strength of its fluctuations.\\

We are grateful to G. Horowitz, M. Roberts and Chris Herzog for useful discussions. S.F. acknowledges the support of the National Science Foundation under Grant No. PHY05-51164. A.M.G. acknowledges financial support from the FEDER
and the Spanish DGI through Project No.
FIS2007-62238. D.R-G acknowledges support from the European Comission through the Marie Curie OIF Action MOIF-CT-2006-38381

\end{document}